\documentclass[prl,twocolumn]{revtex4-1}
\usepackage[latin1]{inputenc}
\usepackage{graphicx}
\usepackage{amssymb}
\usepackage{amsmath}
\usepackage{xspace}
\usepackage{dcolumn}
\usepackage{bm}
\usepackage{color}
\usepackage{float}
\usepackage[extra]{tipa}
\usepackage{mathtools}

\setcitestyle{square,numbers,compress}

\newfont{\tensy}{cmsy10}

\renewcommand{\Re}[0]{\text{Re}\,}
\newcommand{\ie}[0]{i.e.\@\xspace}

\newcommand{\bk}{\boldsymbol{k}}

\newcommand{\bq}{\boldsymbol{q}}
\newcommand{\br}{\boldsymbol{r}}

\newcommand{\rmi}{\text{i}}
\newcommand{\UP}[0]{\uparrow}
\newcommand{\DO}[0]{\downarrow}

\newcommand{\oQ}{\hat{Q}}
\newcommand{\oH}{\hat{H}}

\newcommand{\on}{\hat{n}}

\newcommand{\os}{\hat{s}}

\newcommand{\ZII}{Z$_2$\xspace}

\newcommand{\bit}{\begin{itemize}}
\newcommand{\eit}{\end{itemize}}

\newcommand{\oh}{\mbox{$\frac{1}{2}$}}

\newcommand{\om}[0]{\omega}

\newcommand{\kB}{k_\text{B}}
\newcommand{\nag}{{\phantom{\dag}}}

\newcommand{\las}[0]{\langle}
\newcommand{\ras}[0]{\rangle}

\newcommand{\la}[0]{\left\las}
\newcommand{\ra}[0]{\right\ras}
\newcommand{\ket}[1]{\left|#1\ra}
\newcommand{\bra}[1]{\la#1\right|}

\newcommand{\tr}[0]{\text{tr}}

\begin{document}


\title{Fractionalized Metal in a Falicov-Kimball Model}

\author{Martin Hohenadler}

\author{Fakher F. Assaad}

\affiliation{\mbox{Institut f\"ur Theoretische Physik und Astrophysik,
    Universit\"at W\"urzburg, 97074 W\"urzburg, Germany}}

\begin{abstract}
Quantum Monte Carlo simulations reveal an exotic metallic phase with a
single-particle gap but gapless spin and charge
excitations and a nonsaturating resistivity in a two-dimensional SU(2)
Falicov-Kimball model. An exact duality between this model and an
unconstrained slave-spin theory leads to a classification of the phase
as a fractionalized or orthogonal metal whose low-energy excitations 
have different quantum numbers than the original electrons. Whereas
the fractionalized metal corresponds to the regime of disordered slave
spins, the regime of ordered slave spins is a Fermi liquid. At a critical temperature,
an Ising phase transition to a spontaneously generated 
constrained slave-spin theory of the Hubbard model is observed.
\end{abstract}

\date{\today}

\maketitle

The fractionalization of electrons into objects with new quantum numbers
is among the most fascinating consequences of strong interactions. It is ubiquitous in
one-dimensional (1D) metals, where Fermi liquid theory breaks down
completely and the low-energy properties are instead determined by
collective charge and spin excitations \cite{Giamarchi}. Fractionalization is
less common but physically even richer in higher dimensions, where it
involves emergent degrees of freedom such as spinons or gauge fields
\cite{FradkinBook}. A prime example  is a genuine Mott insulator without
magnetic order that can be classified as a topologically ordered quantum spin
liquid \cite{Hasting04,RevModPhys.89.025003}. Experiments on, for example,
high-temperature superconductors also reveal strange metallic states at
higher temperatures such as non-Fermi liquids \cite{schofield1999non} or bad metals
\cite{PhysRevLett.74.3253}, which 
are believed to be strongly tied to the exotic low-temperature
physics. In orthogonal metals \cite{PhysRevB.86.045128}, with Fermi-liquid-like
transport and thermodynamics but no quasiparticles, non-Fermi-liquid physics
arises from fractionalization and reconciles the absence of quasiparticles in
photoemission with a Fermi surface
according to quantum oscillation measurements \cite{PhysRevB.86.045128}. 
Finally, unusual metallic states have become a focus of applications of the
gauge-gravity duality \cite{faulkner2010strange,PhysRevLett.105.151602,arXiv:1804.04130}.

Recent insights into fractionalized
phases have in particular come from exactly solvable models
\cite{Kitaev20062,PhysRevB.86.045128,zhong2012correlated,arXiv:1804.04130} and designer
Hamiltonians suitable for quantum Monte Carlo (QMC) simulations \cite{PhysRevX.6.041049,gazit2017emergent,arXiv:1804.01095}.
However, the corresponding models have only limited overlap with the standard
models of condensed matter theory. Among the latter, the Hubbard model
\cite{hubbard1963electron} continues to attract interest 
\cite{zheng2017stripe,huang2017numerical,mazurenko2016experimental}, in
significant part due to its expected relevance for high-temperature
superconductivity. The Falicov-Kimball model (FKM)
\cite{falicov1969simple} is much simpler because electrons of one
spin sector remain localized  \cite{hubbard1963electron}. It admits an exact
solution in infinite dimensions where it exhibits a quantum phase transition
\cite{RevModPhys.75.1333}, as well as exact mathematical theorems \cite{gruber1996falicov}.
FKMs are also instrumental to understand correlated electrons out of
equilibrium \cite{RevModPhys.86.779}. While traditionally not
associated with the intricate physics of fractionalization, they have
recently emerged in the context of lattice gauge theories
\cite{PhysRevB.96.205104,smith2018dynamical}. Finally,
FKMs of spinless fermions have been shown to exhibit localization
without disorder \cite{PhysRevLett.117.146601,PhysRevLett.118.266601}.

In this Letter, we show that a fractionalized metallic phase
emerges in QMC simulations of a simple 2D FKM. This
model has no exact solution, but instead reduces to the Hubbard model at
$T=0$ and/or in infinite dimensions. The observed physics can be understood 
by exploiting an exact relation to an unconstrained Ising lattice gauge
theory via a slave-spin representation. Mean-field arguments reveal the basic
mechanism for fractionalization, whereas our simulations
fully account for quantum and thermal fluctuations to establish the
existence of such a phase in this model.

{\it Model.}---We consider the Hamiltonian
\begin{equation}\label{eq:HcQ}
  \hat{H}
  = 
  -t \sum_{\las ij\ras\sigma} ({c}^\dag_{i\sigma} {c}^\nag_{j\sigma} + \text{H.c.} )    
    - {U}\sum_{i} \oQ_i \prod_{\sigma} (\on_{i\sigma}-\oh) \,.
\end{equation}
Here, $c^\dag_{i\sigma}$ creates a spin-$\sigma$ electron at site $i$ of a 
square lattice and $\on^{}_{i\sigma}=c^\dag_{i\sigma}c^\nag_{i\sigma}$. The first term describes
nearest-neighbor hopping. Restricting $\sigma$ to a single value yields a standard
FKM \cite{falicov1969simple} with the localized fermions expressed in terms of the
Ising degrees of freedom $\oQ_i=\pm1$ via the relation $\on^\text{loc}_i=(\oQ_i-1)/2$.
For two flavors $\sigma=\UP,\DO$, the second term in Eq.~(\ref{eq:HcQ})
becomes a three-body interaction of the Hubbard-Ising form $U\sum_i \oQ_i
(\on_{i\UP}-\oh)(\on_{i\DO}-\oh)$.  Generalizations to SU($N$) fermions with
$N>2$ flavors or higher-spin $\oQ_i$ variables are also conceivable. For
$N>1$, the product over  flavors renders Eq.~(\ref{eq:HcQ}) not exactly
solvable even in infinite dimensions; we consider $N=2$ in the following. The
Ising variables $\oQ_i$ are locally conserved, $[\oH,\oQ_i]=0$.
At the particle-hole symmetric point investigated here, Hamiltonian~(\ref{eq:HcQ})
has an O(4)=SO(4)$\times$\ZII symmetry. The SO(4) symmetry is the same as for
the Hubbard model \cite{yang1990so}. The global \ZII symmetry reflects
invariance under $\oQ_i\rightarrow -\oQ_i$ in combination with a
particle-hole transformation \cite{shiba1972thermodynamic} that yields
$U\to-U$; it can be broken at $T>0$ in the 2D case considered.

We use units in which $\kB=t=1$ and consider periodic $L\times L$
lattices. Simulations were done using the
auxiliary-field QMC method \cite{Blankenbecler81} from the Algorithms for
Lattice Fermions library \cite{ALF17}, see also the SM~\cite{SM}.

\begin{figure}[t]
  \includegraphics[width=0.5\textwidth]{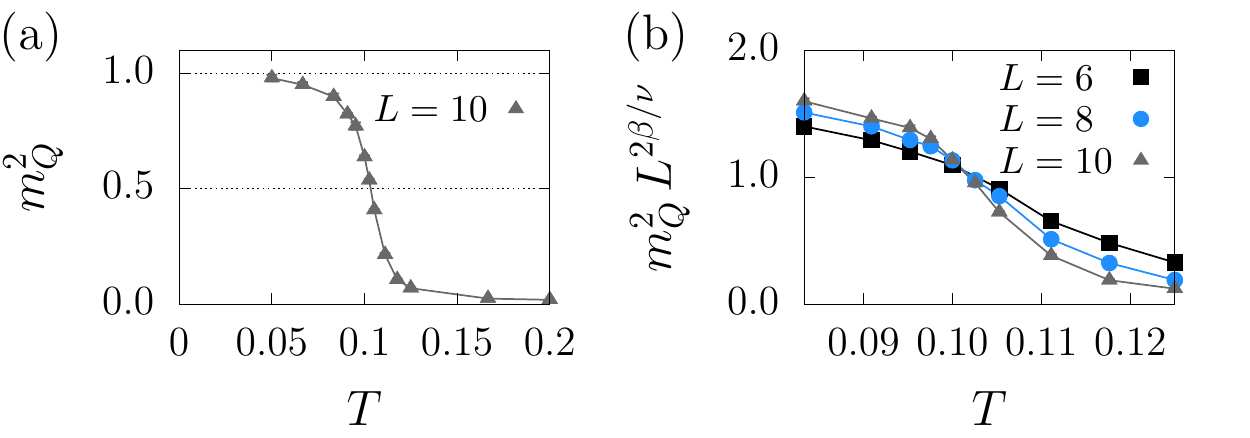}  
  \includegraphics[width=0.5\textwidth]{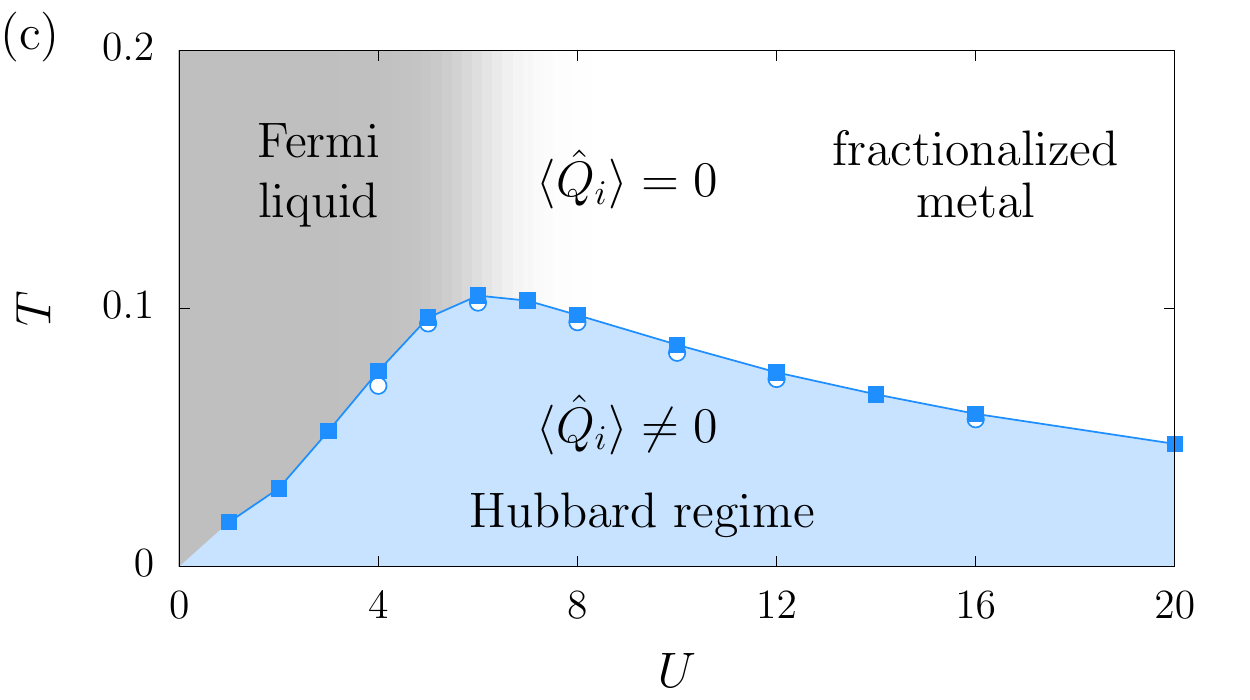}  
  \caption{\label{fig:phasediagram} 
    (a,b) Ising phase transition for $U=6$  and (c) phase diagram of the
    FKM~(\ref{eq:HcQ}) from QMC simulations. The Ising variables
    $\oQ_i$ order ferromagnetically at a critical temperature
    $T_Q$, as revealed by (a) the squared magnetization $m^2_Q$ and (b) the
    finite-size scaling with 2D Ising critical exponents. (c) Phase diagram
    with a high-temperature phase where $\las \oQ_i\ras
    =0$ and a low temperature phase where $\las\oQ_i\ras\neq 0$. The
    disordered phase at $T>T_Q$ consists of two regimes. The Fermi
    liquid for $U\lesssim 4$ and the fractionalized metal for $U\gtrsim 4$ are
    connected by a continuous crossover qualitatively indicated by the color gradient.
    $T_{Q}$ was estimated from $m^2_Q(T_{Q})=0.5$ using
    $L=8$ (solid symbols) and $L=12$ (open symbols), respectively.}
\end{figure}

{\it Ising phase transition.}---Similar to other FKMs \cite{RevModPhys.75.1333}, we find a
finite-temperature phase transition. In our model, the latter corresponds to
a ferromagnetic phase transition of the Ising variables $\oQ_i$ at a critical
temperature $T_Q$ that reduces the symmetry from O(4) to SO(4). Its origin
can be traced back to an exchange coupling $J\sum_{ij} \oQ_i \oQ_j$---mediated by
the itinerant fermions---that is allowed by the symmetries of
Eq.~(\ref{eq:HcQ}) and hence generated. The onset of order is visible from
the squared magnetization per site $m^2_Q = M^2_Q/L^2$, where $M^2_Q =
\frac{1}{L^2}\sum_{ij}\las \oQ_i \oQ_j\ras$, shown in Fig.~\ref{fig:phasediagram}(a). 
The 2D Ising universality is revealed by the finite-size scaling in
Fig.~\ref{fig:phasediagram}(b) with exponents $\beta=1/8$ and $\nu=1$. 
For the phase diagram in Fig.~\ref{fig:phasediagram}(c), we estimated the
critical temperature from $m^2_Q(T_{Q})=0.5$ using $L=8$ and $L=12$.
The dependence of $T_Q$ on $U$ is reminiscent of $T_c$ for the
charge-density-wave (CDW)
transition of the spinless, half-filled FKM \cite{PhysRevB.74.035109,PhysRevB.68.153102}.
In particular, $T_Q=0$ at $U=0$ due to the absence of exchange interactions, and
$T_Q\to0$ for $U\to\infty$ because $T_Q\sim J\sim t^2/U$.
  
Upon replacing the Ising variables $\oQ_i$ by mean-field values $\las \oQ_i\ras
=0$ (for $T>T_Q$) or $\las \oQ_i\ras=m_Q$ (for $T<T_Q$), 
the SU(2) FKM of Eq.~(\ref{eq:HcQ}) reduces to free fermions ($T>T_Q$) or a
Hubbard model ($T<T_Q$). We have verified that below $T_Q$ we quantitatively
recover Hubbard model results for $T\to 0$ \cite{Z2long}, namely an
antiferromagnetic Mott insulator ($m_Q=-1$) or coexisting CDW order and
$s$-wave superconductivity  ($m_Q=+1$), respectively \cite{PhysRevLett.62.1407}.

\begin{figure}[t]
  \includegraphics[width=0.5\textwidth]{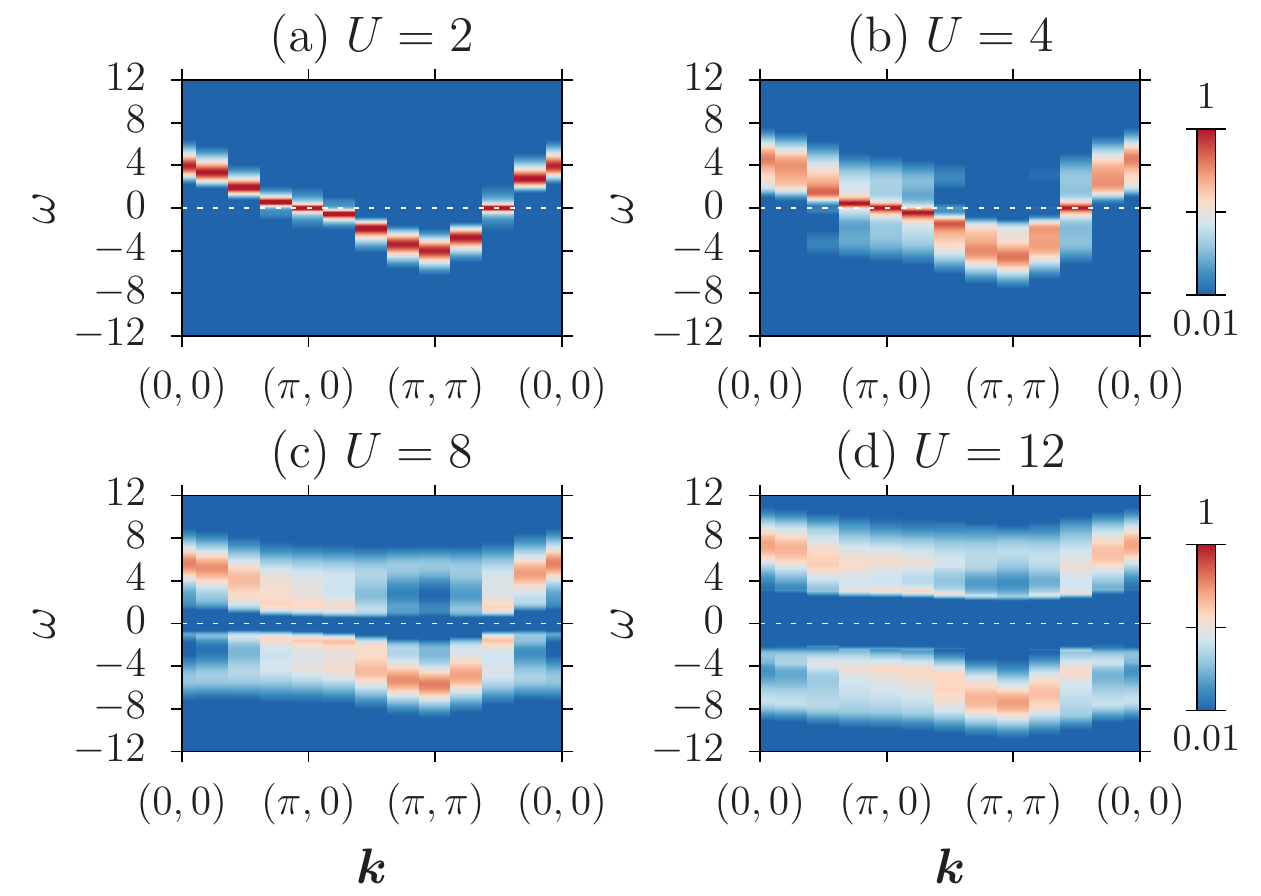} 
  \includegraphics[width=0.5\textwidth]{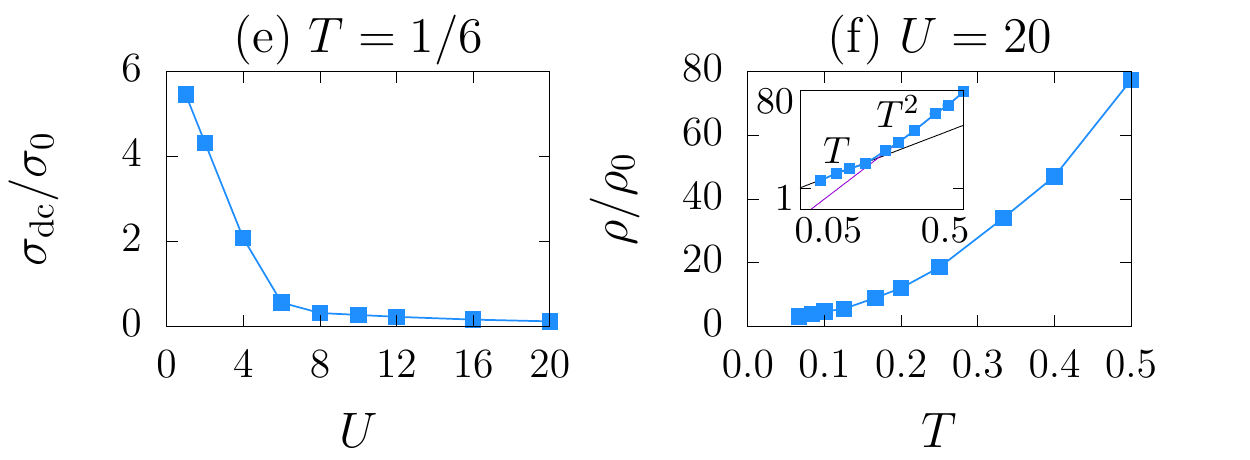}
  \caption{\label{fig:akw-z2} (a--d) Single-particle spectral function
    $A(\bk,\omega)$ at temperature $T={1}/{6}$. (e) Conductivity and (f) resistivity 
    $\rho=1/\sigma_\text{dc}$ (inset: logarithmic scales). Here, $L=8$.}  
\end{figure}

{\it Two distinct metallic regimes.}---The novel physics of this Letter occurs
at $T>T_Q$, where we find two distinct metallic regimes. A mean-field
solution of Eq.~(\ref{eq:HcQ}) with $\las \oQ_i\ras=0$ accounts for the Fermi
liquid observed at weak $U$. The fractionalized metal at large $U$ will
naturally emerge from a slave-spin mean-field theory below.
The two different metallic regimes indicated in Fig.~\ref{fig:phasediagram}(c)
are revealed by the QMC results in Fig.~\ref{fig:akw-z2}. The spin-averaged single-particle
spectral function $A(\bk,\omega)=-\pi^{-1}\mathrm{Im}\, G(\bk,\omega)$
calculated from the Green functions $G_\sigma(\bk,\tau)= \las
c^\dag_{\bk\sigma}(\tau)  c^\nag_{\bk\sigma}(0)\ras$ via analytic
continuation \cite{SM} exhibits coherent, gapless excitations in the Fermi-liquid regime
at $U=2$ [Fig.~\ref{fig:akw-z2}(a)]. For $U\approx 4$, we observe the opening
of a gap that grows with increasing $U$. At $U=12$, the spectrum 
exhibits a large gap and significant broadening [Fig.~\ref{fig:akw-z2}(d)],
\ie, no signatures of Landau quasiparticles. According to
Fig.~\ref{fig:akw-z2}(e), the conductivity $\sigma_\text{dc}$ \cite{SM}
decreases sharply in the Fermi liquid before saturating at a nonzero value
in the fractionalized metal. Finite-size scaling is consistent with
$\sigma_\text{dc}>0$ for $L\to\infty$; moreover, in the fractionalized regime,
$\sigma_\text{dc}$ increases for $T\to 0$ whereas the single-particle
spectral weight at $\omega=0$ decreases strongly \cite{SM}.
Even for $U=20$, we find metallic behavior in terms of a resistivity
$\rho=1/\sigma_\text{dc}$ that increases without saturation with increasing
$T$ [Fig.~\ref{fig:akw-z2}(f);  the inset suggests a crossover from $\rho\sim
T$ to $\rho\sim T^2$]. 

\begin{figure}[t] 
\includegraphics[width=0.25\textwidth]{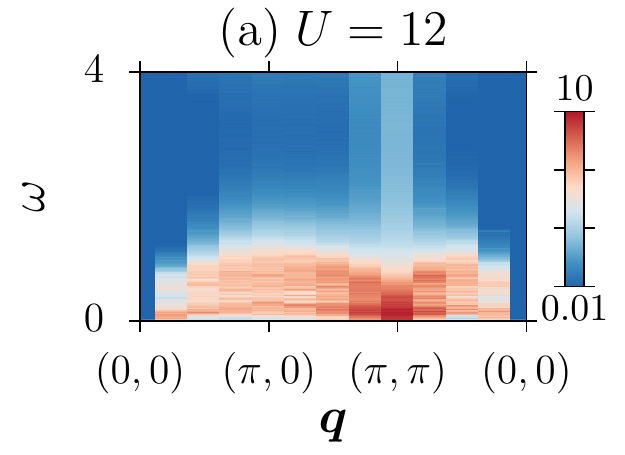}%
\includegraphics[width=0.25\textwidth]{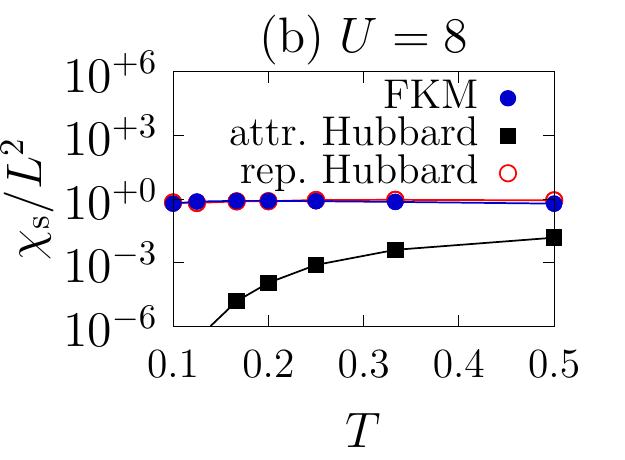}%
 \caption{\label{fig:chis}  
 (a) Dynamic spin structure factor and (b) spin susceptibility 
 of the FKM for $T=1/6$ and $L=8$. (b) also shows results for the
 attractive and the repulsive Hubbard model.}
\end{figure}

Fermi liquid theory cannot reconcile an apparent single-particle gap
\cite{SM} with metallic behavior. However, these features do co-occur in the
pseudogap phase of the attractive Hubbard model at $T_c<T<T^*$ where electrons are bound into uncondensed
singlets ($T_c=0$ for superconductivity at half filling) \cite{PhysRevLett.69.2001}.
In contrast to such a paired Fermi liquid, the fractionalized metal has strongly
renormalized but gapless long-wavelength (\ie, $\bm{q}\to0$) spin
excitations, as visible from the dynamic spin structure factor
$S^z(\bm{q},\omega)$ \cite{SM} in
Fig.~\ref{fig:chis}(a). These excitations give rise to a substantial spin
susceptibility $\chi_\text{s}={\beta}(\las \hat{M}^2\ras -
\las\hat{M}\ras^2)$ (here, $\hat{M}= \sum_i \hat{S}^z_i$) down to
$T_Q\sim t/U^2$, see Fig.~\ref{fig:chis}(b); this behavior is again beyond
Fermi liquid theory where a single-particle gap implies $\chi_\text{s}\to0$ for $T\to0$.
The results for the attractive Hubbard model in
Fig.~\ref{fig:chis}(b) instead exhibit an exponential suppression of
$\chi_\text{s}$ below $T^*\sim U$ \cite{PhysRevB.50.635}. Because of the O(4)
symmetry at half filling, the spin structure factor and the spin
susceptibility of the FKM are identical to their charge counterparts. Hence,
Fig.~\ref{fig:chis} also suggests the existence of gapless charge excitations
and hence a metallic state.  Finally, the repulsive
Hubbard model has a single-particle gap and gapless spin excitations
[Fig.~\ref{fig:chis}(b)] but an exponentially suppressed charge
susceptibility [identical to $\chi_\text{s}$ of the attractive model in
Fig.~\ref{fig:chis}(b)]. In contrast to Fig.~\ref{fig:akw-z2}(f),
its resistivity decreases with increasing $T$, corresponding to insulating
behavior \cite{SM}. Our data support a fractionalized metal that combines the
metallic behavior of the attractive Hubbard model with the gapless spin
excitations of the repulsive Hubbard model to yield a non-Fermi-liquid,
fermionic metal.

{\it Duality and Fractionalization.}---To connect the distinct properties observed in the
metallic regime at large $U$ to fractionalization, we exploit  a duality transformation 
between the FKM~(\ref{eq:HcQ}) and an unconstrained \ZII slave-spin theory. 
To arrive at the latter, we first relabel the states of the local Hilbert space
from
$\{\ket{0}_i,\ket{\UP}_i,\ket{\DO}_i,\ket{\UP\DO}_i\}_c\otimes\{\ket{+1}_i,\ket{-1}_i\}_Q$ to
 $\{\ket{0}_i,\ket{\UP}_i,\ket{\DO}_i,\ket{\UP\DO}_i\}_f\otimes\{\ket{\UP}_i,\ket{\DO}_i\}_s$.
Next, we make the replacements \cite{PhysRevB.81.155118}
\begin{equation}
c^{(\dag)}_{i\sigma}  \mapsto  f^{(\dag)}_{i\sigma} \os^z_i,\quad
  \oQ_i \mapsto \os^x_i (-1)^{\sum_{\sigma} f^{\dagger}_{i\sigma}  f^{\phantom\dagger}_{i\sigma}  }\,.
\end{equation}
Here, $f^{(\dag)}_{i\sigma}$  is a fermionic operator and $\os^z_i,  \os^x_i$
correspond to Pauli spin matrices.  Using the operator identity
$(-1)^{\sum_{\sigma} f^{\dagger}_{i\sigma}  f^{\phantom\dagger}_{i\sigma}  }
\equiv (2\on_{i\UP}-1) (2\on_{i\DO}-1)$ yields the slave-spin formulation of
the FKM~(\ref{eq:HcQ}),
\begin{align}\label{eq:Z2hubbard2}
\hat{H}^{fs} 
= 
-t \sum_{\las ij\ras\sigma} 
( f^\dag_{i\sigma} f^\nag_{j\sigma} \os^z_i \os^z_j +
  \text{H.c.} 
  ) 
 - \frac{U}{4} \sum_i \os^x_i \,.
\end{align}

Equation~(\ref{eq:Z2hubbard2}) locally conserves the $\oQ_i$, $[\oH^{fs},\oQ_i]=0$,
and corresponds to an unconstrained gauge theory in the sense that we do not
impose the Gauss law corresponding to $\oQ_i\ket{\psi}=\ket{\psi}$ or simply $\oQ_i=1$. 
This unconstrained theory is an exact slave-spin representation of Eq.~(\ref{eq:HcQ}).
Enforcing $\oQ_i = 1$ amounts to projecting onto the 4D local Hilbert space
of the Hubbard model and promotes Eq.~(\ref{eq:Z2hubbard2}) to an exact (constrained)
\ZII slave-spin theory of the latter. This also becomes apparent from
Eq.~(\ref{eq:HcQ}) upon setting $\oQ_i=1$. An intriguing question iss
under what conditions the constrained and unconstrained theories are equivalent. According to
Fig.~\ref{fig:phasediagram}, the constraints
$\oQ_i$ are spontaneously generated in the ferromagnetic phase at $T<T_Q$
so that for $T\to0$ the unconstrained theory~(\ref{eq:Z2hubbard2}) becomes an
exact slave-spin representation of the Hubbard model.
Moreover, the constraints are completely irrelevant at $U=0$ [where
both Eq.~(\ref{eq:HcQ}) and Eq.~(\ref{eq:Z2hubbard2}) reduce to 
free fermions] and in infinite
dimensions for any $U$ and $T$; the latter statement holds only at the
particle-hole symmetric point and was previously proved in
the slave-spin representation \cite{schiro2011quantum}. It also
follows directly for the half-filled FKM~(\ref{eq:HcQ}) because the only
nonzero contributions in a diagrammatic expansion in the interaction $U\sum_i
\oQ_i (\on_{i\UP}-\oh)(\on_{i\DO}-\oh)$ contain even numbers of vertices at a
single site (the free propagator is local for $D=\infty$
\cite{khurana1990electrical,muller1989correlated}) and $(\oQ_i)^{2n}=1$.

A mean-field theory of the dual slave-spin model~(\ref{eq:Z2hubbard2})
captures the metallic state observed at strong coupling and relates it to
fractionalization. The product ansatz $\ket{\Phi}_\text{MF} =
\ket{\phi}_f\otimes\ket{\phi}_s$ for the ground-state
decouples the problem into a free-fermion part
$\hat{H}_\text{MF}^{f}  = 
  -{t} \sum_{\las ij\ras\sigma} g_{ij}
  ( f^\dag_{i\sigma} f^\nag_{j\sigma}  +
    f^\dag_{j\sigma} f^\nag_{i\sigma})$
and a transverse-field Ising model
$\hat{H}_\text{MF}^{s} 
  = 
  - t \sum_{\las ij\ras} J_{ij} 
  \os^z_i \os^z_j     
  -\frac{U}{4}\sum_i \os^x_i$
connected by the self-consistency conditions
  $g_{ij} = \las \os^z_i \os^z_j\ras_s$ and 
  $J_{ij} = \sum_\sigma \las f^\dag_{i\sigma} f^\nag_{j\sigma}  + \text{H.c.}
  \ras_f$ \cite{PhysRevB.81.155118}.
The slave spins will be ferromagnetically ordered for $U<U_c$, and disordered
for $U>U_c$. The effect of this transition on the original electrons becomes
clear from their spectral function, $A(\bk,\omega) = \las \os^z_i \ras^2 \delta(\omega-E_{\bk})$
\cite{PhysRevB.86.045128}, where $E_{\bk}$ is the $f$-fermion dispersion. Clearly,
$\las \os^z_i \ras^2$ is directly related to the quasiparticle residue $Z$,
which is finite for $U<U_c$ but vanishes for $U>U_c$. Within single-site mean-field
theories, including dynamical mean-field theory, this
transition is associated with a Mott metal-insulator transition for which
$\las \os^z_i\ras$ serves as an order parameter
\cite{PhysRevB.81.155118,PhysRevB.91.245130}. Beyond single-site mean-field
theories, $\las \os^z_i \os^z_j \ras\neq \las \os^z_i \ras^2$, and the
disordered phase is an orthogonal metal with Drude weight $D\sim \las
\os^z_i \os^z_j \ras$ rather than a Mott insulator \cite{PhysRevB.86.045128}. 

In the context of slave-spin representations, fractionalization amounts
to the dissociation of the physical $c$-electrons into auxiliary
$f$-fermions, which carry the physical U(1) charge \cite{PhysRevB.86.045128}, and
the slave spins $\os^z_i$. Whereas the $c$-fermions are invariant under local
gauge transformations generated by the $\oQ_i$, the $f$-fermions and slave spins
each carry a \ZII gauge charge that manifests itself as $\oQ_i
f^{(\dag)}_{i\sigma} \oQ_i  = - f^{(\dag)}_{i\sigma} $,  $\oQ_i \os^z_i \oQ_i
= - \os^z_i$. While this charge is
strictly conserved only in constrained gauge theories,
the notion of fractionalization remains meaningful in a broader context,
including mean-field theories, where the constraints are either ignored or
imposed on average \cite{PhysRevB.81.155118}, and unconstrained gauge
theories such as Eq.~(\ref{eq:Z2hubbard2}), where the charge is conserved in
space but not in time. In particular, the orthogonal metal emerging in
mean-field theory at $U>U_c$ from the disordering of the slave spins
may be regarded as fractionalized in the sense that the metallic properties
are carried by the \ZII-charged $f$-fermions that are orthogonal
\cite{PhysRevB.86.045128} to the gauge-invariant $c$-fermions.

\begin{figure}[t] 
\includegraphics[width=0.5\textwidth]{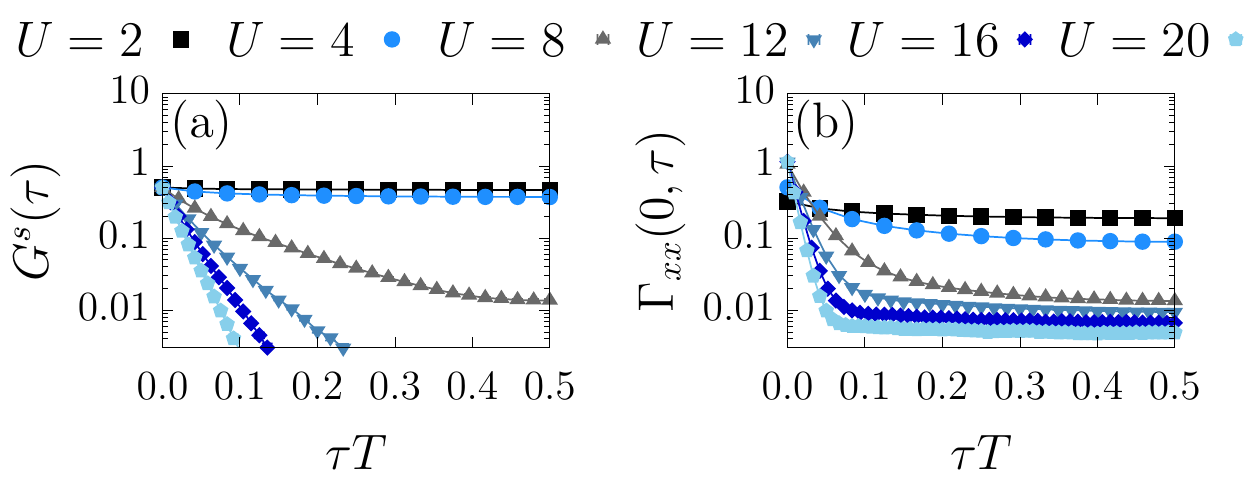}
 \caption{\label{fig:crossover} 
 (a) Slave-spin Green function $G^s(\tau)$ and (b) current-current correlator
 $\Gamma_{xx}(\bq=0,\tau)$ for $L=8$ and $T=1/6$. 
 }
\end{figure}

The mean-field fractionalization scenario is essentially
borne out by our QMC results for the
FKM: as shown in Fig.~\ref{fig:akw-z2}, the single-particle spectrum has a
gap at large $U$ but the system remains metallic. 
Within our unbiased QMC  approach,
the mean-field phase transition of the slave spins is replaced by an
order-disorder crossover reflected in the slave-spin correlator $G^s(\tau)= \las
\os^z_{i}(\tau)  \os^z_{i}\ras$ in Fig.~\ref{fig:crossover}(a) and directly
related to the opening of the single-particle gap in Fig.~\ref{fig:akw-z2}. The disorder of the slave spins strongly
enhances scattering and suppresses coherent quasiparticle motion
[Fig.~\ref{fig:akw-z2}(e)]. However, the current correlator
$\Gamma_{xx}(\bq=0,\tau)$ \cite{SM} in Fig.~\ref{fig:crossover}(b) 
remains gapless even for large $U$.

{\it Discussion}---While the non-Fermi-liquid regime at large $U$ exists
independent of the slave-spin representation, the
latter reveals the fractionalization and close conceptual relations to
orthogonal metals \cite{PhysRevB.86.045128}.  On the other hand, our
findings differ in a number of important details from previous mean-field and exact
realizations of such states \cite{PhysRevB.86.045128}. 
First, our simulations preserve the local \ZII gauge symmetry of
Eq.~(\ref{eq:Z2hubbard2}), in accordance with Elitzur's theorem
\cite{RevModPhys.51.659}. This symmetry---reflecting invariance under the
local transformation $f^{(\dag)}_{i\sigma}\mapsto - f^{(\dag)}_{i\sigma}$,
$\os^z_i\mapsto -\os^z_i$ generated by $\oQ_i$---implies that the spatial
correlations $\las \os^z_i \os^z_{j\neq i}\ras$ responsible for a nonzero
Drude weight in mean-field theory are zero
\cite{PhysRevX.6.041049}. Accordingly, the slave spins undergo a crossover
[in imaginary time, see Fig.~\ref{fig:crossover}(a)], instead of a phase
transition. Similarly, the $f$-fermions are localized because they also carry
\ZII charge and the gapped but dispersive single-particle excitations in
Fig.~\ref{fig:akw-z2}(d) instead emerge from the combination of
imaginary-time correlations (\ie, quantum
fluctuations) and vertex corrections. If the latter are absent, as in
infinite dimensions, a single-particle gap always implies insulating
behavior \cite{khurana1990electrical}. In this limit, non-Fermi-liquid
behavior can arise without fractionalization from spin freezing \cite{PhysRevLett.101.166405}.
Whereas the orthogonal metals
in the exactly solvable models of Ref.~\cite{PhysRevB.86.045128} are noninteracting, 
transport and thermodynamic properties are strongly renormalized by
interactions in the present, correlated fractional metal. Finally, in
contrast to the $t$-$J$ model with random interaction \cite{arXiv:1804.04130},
our fractionalized phase arises in a fully translation-invariant setting.

Our work has connections to several other areas of current interest. A 1D
unconstrained gauge theory (equivalent to a spinless FKM) was recently shown
to exhibit localization without disorder \cite{PhysRevLett.118.266601}. The
quantum percolation mechanism in the 2D case \cite{smith2018dynamical} may be
connected to the metallic behavior observed here. At high
temperatures, our Falicov-Kimball problem becomes equivalent to a Hubbard
model with an annealed, disordered interaction $U_i=\pm U$. For bosons, a
random Hubbard interaction supports many-body localization \cite{PhysRevA.95.021601}. The slave-spin formulation~(\ref{eq:Z2hubbard2}) provides a link
to recent simulations of lattice gauge theories coupled
to fermions that exhibit exotic phases and phase transitions
\cite{PhysRevX.6.041049,gazit2017emergent,arXiv:1804.01095}, as well as to
Sachdev-Ye-Kitaev models \cite{arXiv:1804.04130}. Progress on cold-atom
realizations of FKMs and Hubbard models
\cite{lewenstein2007ultracold,esslinger2010fermi,mazurenko2016experimental}
as well as lattice gauge theories \cite{zohar2015quantum} even promises the
possibility of experimentally observing the fractionalized metal, facilitated
by its stability at high temperatures.

In summary, we have presented unbiased numerical evidence for a
non-Fermi-liquid phase in a simple 2D Falicov-Kimball model. This
Fermi metal differs from phases of incoherently
paired fermions (\ie, bosons) such as the paired Fermi liquid known from the
attractive Hubbard model, and previous realizations of orthogonal metals.
The exact relation to an unconstrained slave-spin representation
allowed us to understand the physics in terms of fractionalization of the
original electrons.

\begin{acknowledgments}
We thank M. Fabrizio and A. Honecker for helpful discussions. This work was
supported by the DFG through SFB 1170 ToCoTronics. We thank the John von
Neumann Institute for Computing (NIC) for computer resources on the
JURECA~\cite{Juelich} machine at the J\"ulich Supercomputing Centre (JSC).
\end{acknowledgments}


%

\newpage

\begin{widetext}

\begin{center}
\large{\bf Supplemental Material}
\end{center}

\subsection*{Methods}\label{sec:model}

Given the exact duality that relates Eqs.~(1) and~(4), the simulations
were carried out in the slave-spin representation. The site
Ising variables in Eq.~(4) were mapped to bond
Ising variables. The resulting Hamiltonian takes the form
\begin{align}\label{eq:Z2hubbard3}
  \hat{H}^{fZ}_0 
  = 
  -t \sum_{\las ij\ras\sigma}
  (f^\dag_{i\sigma}f^\nag_{j\sigma} +
  \text{H.c.})\,\hat{Z}_{ij} 
  -
  \frac{U}{4}\sum_i \hat{X}_{i+\hat{x}} \hat{X}_{i-\hat{x}}
\hat{X}_{i+\hat{y}}  \hat{X}_{i-\hat{y}}
\,,
\end{align}
where the second term flips all four bond Ising spins attached to site
$i$. In this representation, the constraint $\prod_{\las
  ij\ras\in\partial\Box}\hat{Z}_{ij} = 1$ ($\partial\Box$ denotes the bonds
of a plaquette) holds and ensures that the number
of degrees of freedom remains constant. This constraint was imposed in the simulations.

Simulations were carried out using the auxiliary-field QMC code
from the Algorithms for Lattice Fermions (ALF) library \cite{ALF17}. In the
present case, the role of the auxiliary fields is taken by the bond Ising
variables in Eq.~(\ref{eq:Z2hubbard3}).
The grand-canonical partition function is written as a Euclidean path
integral over Ising spin configurations $\bm{Z}=\{Z_{i\tau}\}$,
\begin{equation}
  Z = \tr\, e^{-\beta (\hat{H}-\mu \hat{N})} = \int \mathcal{D}[\bm{Z}]\,e^{-S[\bm{Z}]}\,.
\end{equation}
As usual, imaginary time was discretized with a Trotter timestep
$\Delta\tau=\beta/L$ ($\beta=1/T$ is the inverse temperature); we used
$\Delta\tau U\leq 0.1$.
The configuration weight can be written as $e^{-S} = e^{-S_0}\det[1 +
B(\bm{Z})]$. Here, $S_0$ describes the spin dynamics due to the 
transverse field in Eq.~(\ref{eq:Z2hubbard3}), whereas $B$ is a product
over time slices of exponentials of the hopping term that contains the
fermion-spin coupling. Because $S$ is real, there is no sign problem.
Since each bond spin $\hat{Z}_{ij}$ is related to two site spins
$\os^z_i$ and $\os^z_j$, the minimal update consists of
flipping all four bond spins connected to a site $i$. 
To make the mapping between site and bond spins bijective, we
stored a reference eigenvalue $s^z_{i_0}$ for each configuration.

Observables were measured using
the single-particle Green function and Wick's theorem \cite{Assaad08_rev}. Apart
from the gauge-invariant observables defined in terms of the original fermions, we
also measured correlation functions of the slave spins $\os^z_i$. Dynamic
correlation functions are accessible in imaginary time, and can be
analytically continued to real frequencies using the maximum entropy method \cite{Beach04a}.
The results for the Hubbard model were obtained from auxiliary-field
simulations based on an SU(2) symmetric Hubbard-Stratonovich decomposition.

\subsection*{Observables}

Here we provide definitions for some of the observables shown in the Letter.
The fermionic spin operator is defined as $\hat{S}^z_i=(\on_{i\UP}-\on_{i\DO})/2$.
The Lehman representation of the dynamic spin structure factor reads
\begin{equation}
  S^z(\bq,\omega) =  \frac{1}{Z}\sum_{mn} {|\bra{m} \hat{S}^z_q \ket{n}|}^2
  e^{-\beta E_n}   \delta(E_n - E_m -\om)\,,
\end{equation}
where $\ket{m}$ is an eigenstate with eigenvalue $E_m$.
The current-current correlation function $\Gamma_{xx}(\bq,\tau)$ is the 
Fourier-transform of 
\begin{equation}
  \Gamma_{xx}(\br,\tau) = \las \hat{\jmath}_x(\br,\tau) \hat{\jmath}_x(0,0)\ras 
\end{equation}
with the current operator
$\hat{\jmath}_x(\br) = \rmi \sum_\sigma(
 c^\dag_{\br+\hat{e}_x,\sigma}c^\nag_{\br,\sigma}
- c^\dag_{\br,\sigma} c^\nag_{\br+\hat{e}_x,\sigma}
 )$. From $\Gamma_{xx}(\bq,\tau)$, we extracted the dc conductivity
$\sigma_\text{dc}=\lim_{\omega\to0}\Re\sigma_\text{reg}$ without 
analytic continuation by using  \cite{PhysRevB.59.4364}
\begin{equation}\label{eq:sigmafromJJ}
  \sigma_\text{dc}
  \approx
  \frac{\beta^2}{\pi}
  \Gamma_{xx}(\bq=0,\tau=\beta/2)\,.
\end{equation}

\subsection*{Absence of superconductivity}

We have directly verified the absence of any signatures of superconducting
behavior in the FKM by calculating the superfluid density 
 $D_s = - e_{\text{kin},x} - \Gamma_{xx}(q_x=0,q_y\to0,\omega=0)$ where $e_{\text{kin},x}
= \sum_\sigma\las c^\dag_{\boldsymbol{r}\sigma}c^\nag_{\boldsymbol{r}+\hat{e}_x,\sigma} +
\text{H.c.}\ras $ \cite{PhysRevB.47.7995}. It scales to zero for all
parameters considered.

\subsection*{Finite-size scaling of $\sigma_\text{dc}$}

To demonstrate that the nonzero conductivity in the fractionalized metallic phase
is not a finite-size artifact, Fig.~\ref{fig:sigma-fss} shows a finite-size
scaling for different values of $U$. Whereas $\sigma_\text{dc}$ decreases
slightly in the Fermi liquid phase at $U=4$, it actually increases with
increasing $L$ and has a very weak size dependence in the fractionalized phase.

\begin{figure}[t] 
\includegraphics[width=0.45\textwidth]{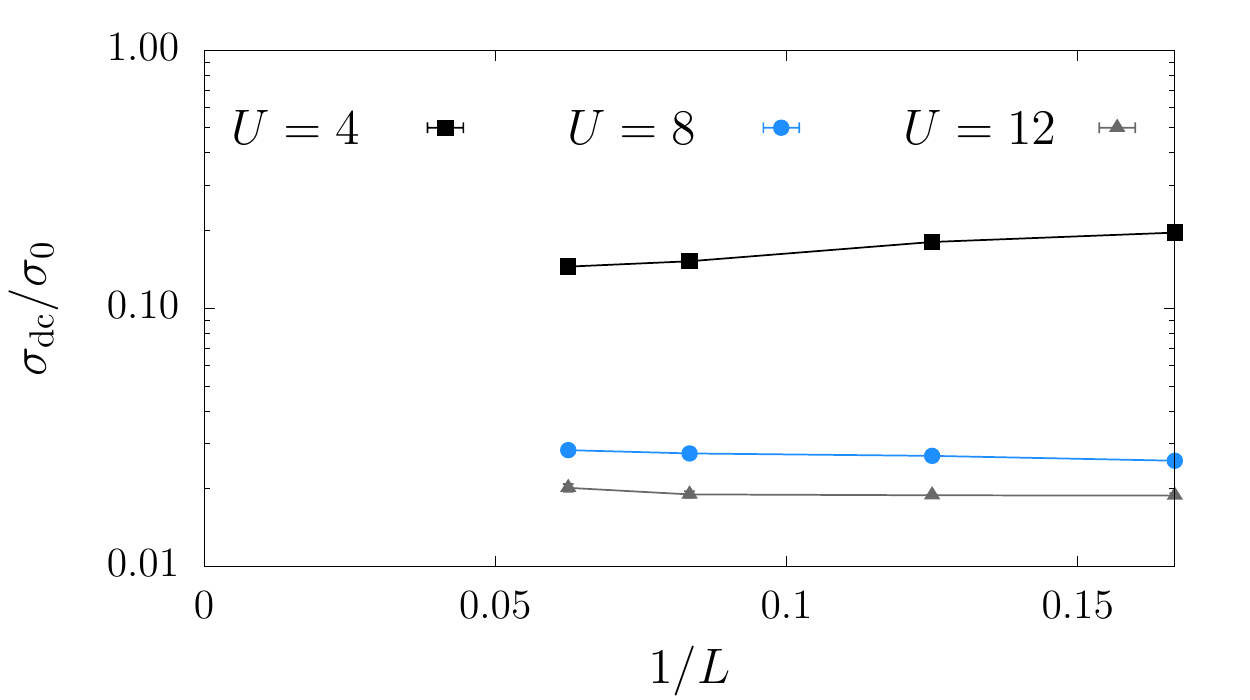}
 \caption{\label{fig:sigma-fss} 
   Finite-size scaling of the conductivity at $T=1/6$.
}
\end{figure}

\subsection*{Single-particle gap}

The opening of a gap in the single-particle spectral function of the
original fermions visible in Fig.~2 can also be inferred directly from
the corresponding imaginary-time Green function $G(\bk,\tau)$. According to
Fig.~\ref{fig:electronicgreenfunction}(a), in contrast to
the Fermi liquid regime represented by $U=2$, $G(\bm{k}_\text{F},\tau)$ 
[with $\bm{k}_\text{F}=(\pi/2,\pi/2)$] at
$U=12$ exhibits an exponential decay that reflects the substantial gap in
$A(\bm{k},\omega)$. 

\begin{figure}[b] 
\includegraphics[width=0.9\textwidth]{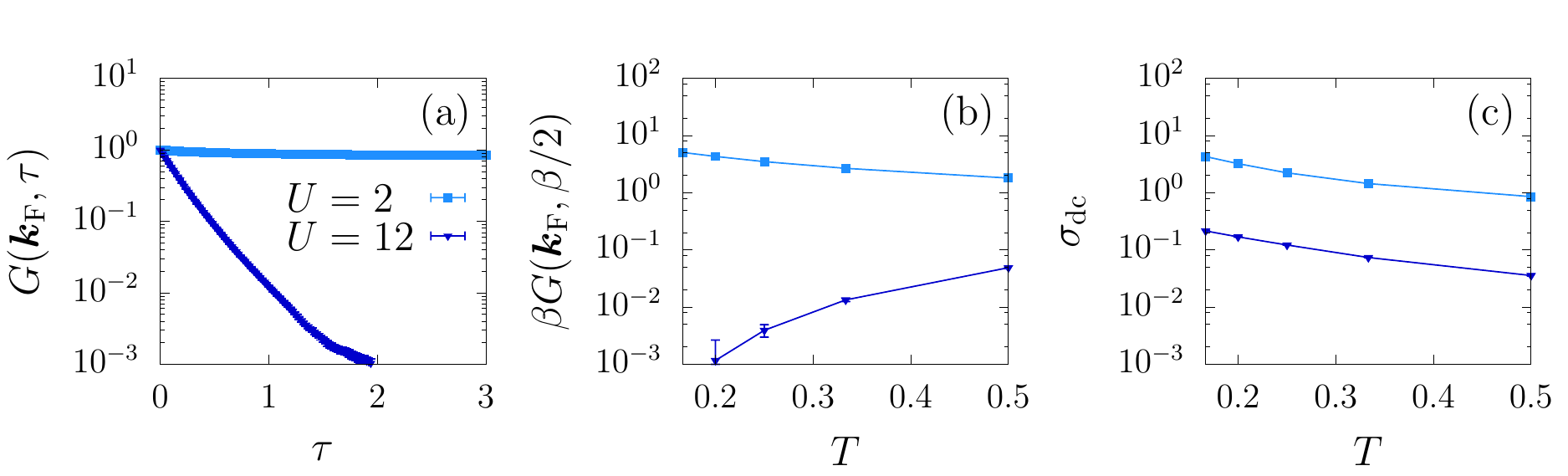}
 \caption{\label{fig:electronicgreenfunction}   
  (a) Electronic Green function of the FKM for
   $\bm{k}_\text{F}=(\pi/2,\pi/2)$ and $T=1/6$, (b) rescaled Green function $\beta
   G(\bm{k}_\text{F},\beta/2)$ according to Eq.~(\ref{eq:qpfromG}), (c)
   conductivity from Eq.~(\ref{eq:sigmafromJJ}). Here, $L=8$. 
 }
\end{figure}

We can relate the value $G(\bm{k}_\text{F},\beta/2)$ to the spectral weight at
the Fermi level by using 
\begin{equation}\label{eq:qpfromG}
  A(\bk,\omega=0)
  =
  \lim_{\beta\to\infty}
  \beta G(\bk,\tau=\beta/2)\,.
\end{equation}
As shown in Fig.~\ref{fig:electronicgreenfunction}(b), $\beta G(\bm{k}_\text{F},\beta/2)$
increases with decreasing temperature in the Fermi liquid regime and
approaches a nonzero value at $T=0$ consistent with gapless excitations and
hence quasiparticles. In contrast, in the fractionalized metal at $U=12$, 
$\beta G(\bm{k}_\text{F},\beta/2)$ appears to scale to zero, consistent with a gap.
On the other hand, in both regimes, the conductivity
increases with decreasing temperature, as expected for metallic states (see below).

\subsection*{Conductivity of the Hubbard model}

The usual definition of metallic behavior at finite temperature is for
the conductivity to decrease with increasing temperature. In contrast,
insulating behavior amounts to a thermally activated conductivity, \ie, an
increase of the conductivity with increasing temperature. For the resistivity
shown in Fig.~2(d), the situation is exactly reversed.

Figure~\ref{fig:conductivity}(a) shows that the conductivity increases with
increasing temperature in the attractive Hubbard model, similar to the
fractionalized metal. In contrast, the repulsive Hubbard model exhibits
insulating behavior [Fig.~\ref{fig:conductivity}(b)]. 

\begin{figure}[t] 
\includegraphics[width=0.6\textwidth]{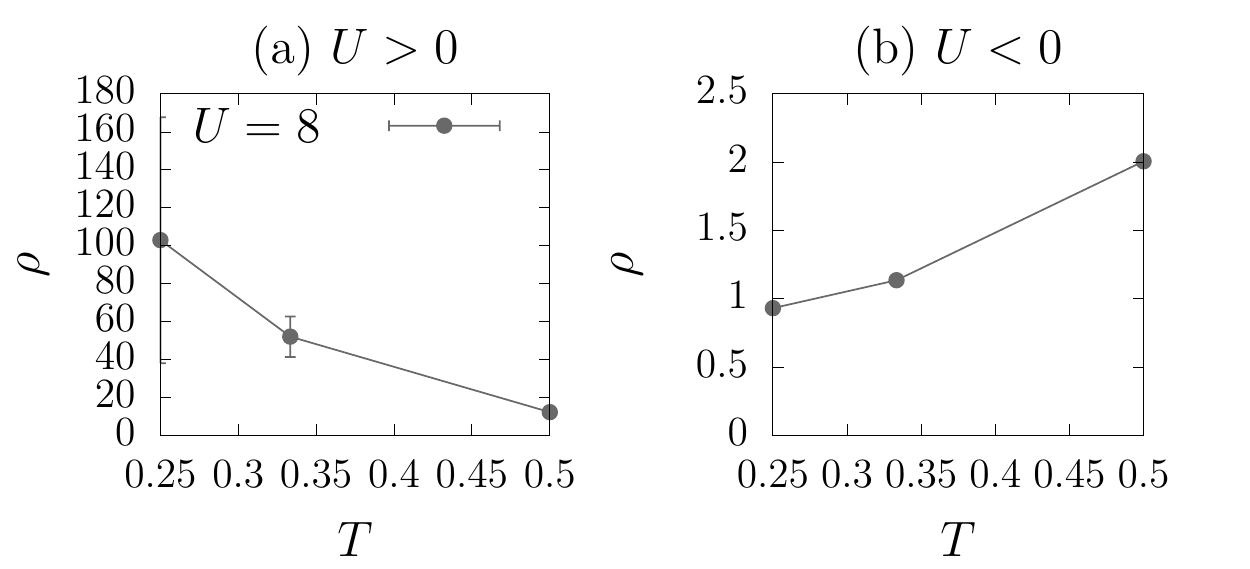}
 \caption{\label{fig:conductivity} 
 Conductivity of (a) the attractive and (b) the repulsive Hubbard model.
 Here, $U=\pm 8$ and $L=8$.
}
\end{figure}

\end{widetext}

\end{document}